\documentclass[preprint,5p]{elsarticle}

\usepackage{graphicx}                       
\usepackage{array}  	
\usepackage{amsmath, amssymb}  
\usepackage{float} 		
\usepackage{subfigure} 

\begin{document}

\newcommand{\kl}[1]{\left(#1\right)}
\newcommand{\ilim}[2]{\int\limits_{#1}^{#2}}																
\newcommand{\slim}[2]{\sum\limits_{#1}^{#2}}																
\newcommand{\liml}[2]{\lim\limits_{#1 \rightarrow #2}}											
\newcommand{\sumsig}{\sum\limits_\sigma} 																		
\newcommand{\sumsigS}{\sum\limits_{\sigma'}} 																
\newcommand{\sumksig}{\sum\limits_{\boldsymbol{k}\sigma}}														
\newcommand{\sumk}{\sum\limits_{\boldsymbol{k}}}													 					
\newcommand{\sumisig}{\sum\limits_{i,\sigma}}																
\newcommand{\sumijsig}{\sum\limits_{i,j,\sigma}}														
\newcommand{\sumij}{\sum\limits_{i,j}}																			
\newcommand{\sumi}{\sum\limits_i}																						
\newcommand{\sumj}{\sum\limits_j}																						
\newcommand{\summ}{\sum\limits_m}																						
\newcommand{\Kom}[2]{\left[#1,#2\right]_-}
\newcommand{\AKom}[2]{\left[#1,#2\right]_{+}}
\newcommand{\GFkl}[2]{\left\langle \left\langle #1; #2 \right\rangle\right\rangle_E}
\newcommand{\erw}[1]{\left\langle #1 \right\rangle}
\newcommand{\Hs}{\sumijsig\kl{T_{ij} - \mu\delta_{ij}}c^\dagger_{i\sigma}c_{j\sigma}}
\newcommand{\Hfeins}{\sumisig\kl{E_f - \mu}f^\dagger_{i\sigma}f_{i\sigma}}
\newcommand{\Hfzwei}{\frac{U_f}2\sumisig n_{fi\sigma}n_{fi-\sigma}}
\newcommand{\Hsf}{-\frac{J'}2\sumisig\left\{\kl{\frac12\sumsigS z_\sigma z_{\sigma'}n_{fi\sigma'}n_{si\sigma}} + f^\dagger_{i\sigma}f_{i-\sigma}c^\dagger_{i-\sigma}c_{i\sigma}\right\}}
\newcommand{\HV}{V\sumisig\kl{c^\dagger_{i\sigma}f_{i\sigma} + f^\dagger_{i\sigma}c_{i\sigma}}}
\newcommand{\HNB}{H_{\text{NB}}}
\newcommand{\Hstilde}{\sumijsig\kl{\Tijs - \mu\delta_{ij}}c^\dagger_{i\sigma}c_{j\sigma}}
\newcommand{\Hftilde}{\sumisig\kl{\kl{\Efs - \mu}\nfis + \frac{U_f}2 \nfis\nfims}}
\newcommand{\HVtilde}{\sumisig\Vs\kl{\cisd\fis + \fisd\cis}}
\newcommand{\cis}{c_{i\sigma}}
\newcommand{\cjs}{c_{j\sigma}}
\newcommand{\cims}{c_{i-\sigma}}
\newcommand{\cjms}{c_{j-\sigma}}
\newcommand{\cisd}{c^\dagger_{i\sigma}}
\newcommand{\cjsd}{c^\dagger_{j\sigma}}
\newcommand{\cimsd}{c^\dagger_{i-\sigma}}
\newcommand{\cjmsd}{c^\dagger_{j-\sigma}}
\newcommand{\cksd}{c_{\boldsymbol{k}\sigma}^\dagger}
\newcommand{\cks}{c_{\boldsymbol{k}\sigma}}
\newcommand{\cs}{c_\sigma}
\newcommand{\csd}{c^\dagger_\sigma}
\newcommand{\cms}{c_{-\sigma}}
\newcommand{\cmsd}{c^\dagger_{-\sigma}}
\newcommand{\fis}{f_{i\sigma}}
\newcommand{\fms}{f_{m\sigma}}
\newcommand{\fmsd}{f_{-\sigma}^\dagger}
\newcommand{\fs}{f_\sigma}
\newcommand{\fjs}{f_{j\sigma}}
\newcommand{\fims}{f_{i-\sigma}}
\newcommand{\fjms}{f_{j-\sigma}}
\newcommand{\fisd}{f^\dagger_{i\sigma}}
\newcommand{\fjsd}{f^\dagger_{j\sigma}}
\newcommand{\fimsd}{f^\dagger_{i-\sigma}}
\newcommand{\fjmsd}{f^\dagger_{j-\sigma}}
\newcommand{\ajs}{a_{j\sigma}}
\newcommand{\aisd}{a^\dagger_{i\sigma}}
\newcommand{\ncis}{n_{si\sigma}}
\newcommand{\nisig}{n_{i\sigma}}
\newcommand{\nimsig}{n_{i-\sigma}}
\newcommand{\ncims}{n_{si-\sigma}}
\newcommand{\nfis}{n_{fi\sigma}}
\newcommand{\nfims}{n_{fi-\sigma}}
\newcommand{\ncisS}{n_{si\sigma'}}
\newcommand{\ncimsS}{n_{si-\sigma'}}
\newcommand{\nfisS}{n_{fi\sigma'}}
\newcommand{\nfimsS}{n_{fi-\sigma'}}
\newcommand{\nfs}{n_{f\sigma}}
\newcommand{\ncs}{n_{s\sigma}}
\newcommand{\nfms}{n_{f-\sigma}}
\newcommand{\nms}{n_{-\sigma}}
\newcommand{\ns}{n_\sigma}
\newcommand{\nis}{n_{i\sigma}}
\newcommand{\nims}{n_{i-\sigma}}
\newcommand{\nup}{n_\uparrow}
\newcommand{\ndown}{n_\downarrow}
\newcommand{\nfu}{n_{f\uparrow}}
\newcommand{\nfd}{n_{f\downarrow}}
\newcommand{\ncu}{n_{s\uparrow}}
\newcommand{\ncd}{n_{s\downarrow}}
\newcommand{\gew}{\widehat{\alpha_{i\sigma}}\kl{\boldsymbol{k}}}
\newcommand{\gewONE}{\widehat{\alpha_{1\sigma}}\kl{\boldsymbol{k}}}
\newcommand{\gewTWO}{\widehat{\alpha_{2\sigma}}\kl{\boldsymbol{k}}}
\newcommand{\gewok}{\widehat{\alpha_{i\sigma}}}
\newcommand{\gewONEok}{\widehat{\alpha_{1\sigma}}}
\newcommand{\gewTWOok}{\widehat{\alpha_{2\sigma}}}
\newcommand{\boldsymbolR}{\boldsymbol{R}}											
\newcommand{\boldsymbolS}{\boldsymbol{S}}											
\newcommand{\zs}{z_\sigma}												
\newcommand{\Efs}{E_{f\sigma}}										
\newcommand{\epsk}{\epsilon\kl{\boldsymbol{k}}}						
\newcommand{\epssk}{\epsilon_{\sigma}\kl{\boldsymbol{k}}}	
\newcommand{\Tij}{T_{ij}}													
\newcommand{\Tijs}{T_{ij\sigma}}									
\newcommand{\dij}{\delta_{ij}}													
\newcommand{\gs}{\gamma_\sigma}										
\newcommand{\gms}{\gamma_{-\sigma}}								
\newcommand{\Vs}{V_\sigma}												
\newcommand{\ms}{m_\sigma}												
\newcommand{\PARas}{a_\sigma}											
\newcommand{\PARbs}{b_\sigma}											
\newcommand{\DMS}{\Delta_{-\sigma}}
\newcommand{\DS}{\Delta_{\sigma}}
\newcommand{\XMS}{X_{-\sigma}}
\newcommand{\YMS}{Y_{-\sigma}}
\newcommand{\Tisk}{T_{i\sigma}(\boldsymbol{k})}						
\newcommand{\Tisx}{T_{i\sigma}(x)}								
\newcommand{\Efij}{E_{f(ij)}}
\newcommand{\Efk}{E_f(\boldsymbol{k})}
\newcommand{\Efisk}{E_{f,i\sigma}(\boldsymbol{k})}
\newcommand{\Efisx}{E_{f,i\sigma}(x)}
\newcommand{\EfONEsk}{E_{f,1\sigma}(\boldsymbol{k})}
\newcommand{\EfTWOsk}{E_{f,2\sigma}(\boldsymbol{k})}
\newcommand{\EfONEmsk}{E_{f,1-\sigma}(\boldsymbol{k})}
\newcommand{\EfTWOmsk}{E_{f,2-\sigma}(\boldsymbol{k})}
\newcommand{\cGFem}{\ensuremath{G_{\boldsymbol{k}\sigma}(E)}}
\newcommand{\cGFemN}{\ensuremath{G^{0}_{\boldsymbol{k}\sigma}(E)}}
\newcommand{\fGFem}{\ensuremath{F_{\boldsymbol{k}\sigma}(E)}}
\newcommand{\pGFone}{\ensuremath{P^{(1)}_{\boldsymbol{k}\sigma}(E)}}
\newcommand{\pGFtwo}{\ensuremath{P^{(2)}_{\boldsymbol{k}\sigma}(E)}}
\newcommand{\cGFemVnull}{\ensuremath{G^{V=0}_{\boldsymbol{k}\sigma}(E)}}
\newcommand{\fGFemVnull}{F^{V=0}_\sigma(E)}
\newcommand{\fGFemVnullk}{F^{V=0}_{\boldsymbol{k}\sigma}(E)}

\newcommand{\cGFmfn}{G^{(\text{MFN})}_{\boldsymbol{k}\sigma}(E)}
\newcommand{\cGFmfnn}{G^{(\text{MFN},0)}_{\boldsymbol{k}\sigma}(E)}
\newcommand{\fGFmfn}{F^{(\text{MFN})}_{\boldsymbol{k}\sigma}(E)}
\newcommand{\fGFmfnn}{F^{(\text{MFN},V=0)}_{\boldsymbol{k}\sigma}(E)}
\newcommand{\pGFmfn}{P^{(\text{MFN})}_{\boldsymbol{k}\sigma}(E)}
\newcommand{\pGFmfnn}{P^{(\text{MFN},0)}_{\boldsymbol{k}\sigma}(E)}
\newcommand{\fGFmfnN}{F^{(\text{MF},V=0)}_{\sigma}(E)}
\newcommand{\frGFem}{F^{\text{(r,EM)}}_{\boldsymbol{k}\sigma}(E)}
\newcommand{\fGFemVnullok}{F^{\text{EM,}V_0}_{\sigma}(E)}
\newcommand{\MNks}{M^{(0)}_{\boldsymbol{k}\sigma}}											
\newcommand{\Gonems}{G_{1-\sigma}}
\newcommand{\Gtwoms}{G_{2-\sigma}}
\newcommand{\selfFem}{\Sigma^{\text{(f,EM)}}_{\sigma}(E+\mu)}				
\newcommand{\selfFemKone}{\Sigma^{\text{(1,f,EM)}}_{\boldsymbol{k}\sigma}(E)}	
\newcommand{\selfFemKtwo}{\Sigma^{\text{(2,f,EM)}}_{\boldsymbol{k}\sigma}(E)}		
\newcommand{\selfFemKi}{\Sigma^{\text{(i,f,EM)}}_{\boldsymbol{k}\sigma}(E)}		
\newcommand{\selfFemi}{\Sigma^{\text{(i,f,EM)}}_{\sigma}(E)}		
\newcommand{\SFU}{\Sigma^{(f),\text{U}}_{ij\sigma}(E)}			
\newcommand{\SFUk}{\Sigma^{(f),\text{U}}_{\boldsymbol{k}\sigma}(E)}			
\newcommand{\SF}{\Sigma^{(f),\text{sf}}_{ij\sigma}(E)}			
\newcommand{\SFk}{\Sigma^{(f),\text{sf}}_{\boldsymbol{k}\sigma}(E)}			
\newcommand{\selfCemLok}{\Sigma^{\text{(s,EM)}}_{ij\sigma}(E)}		
\newcommand{\selfCem}{\Sigma^{\text{(s,EM)}}_{\boldsymbol{k}\sigma}(E+\mu)}	
\newcommand{\selfISALok}{\Sigma^{\text{ISA}}_{ii\sigma}(E+\mu)}			
\newcommand{\selfISA}{\Sigma^{\text{(s),ISA1}}_\sigma(E)}			
\newcommand{\selfCmfnLok}{\Sigma^{\text{(s,MFN)}}_{ij\sigma}(E)}	
\newcommand{\SC}{\Sigma^{(s),\text{sf}}_{ij\sigma}(E)}			
\newcommand{\SCk}{\ensuremath{\Sigma^{(s),\text{sf}}_{\boldsymbol{k}\sigma}(E)}}			
\newcommand{\rhoN}{\varrho_0(\widetilde{E})}
\newcommand{\rN}{\varrho_0(x)}

\frontmatter
	\title{Band model for the understanding of ferromagnetism in semiconductors and insulators}

	\author[UHH]{M.~H\"ansel\corref{cor}}
	\ead{mhaensel@physnet.uni-hamburg.de}
	
	\author[BLN]{W.~Nolting}
	\author[UHH]{M.~Potthoff}

	\address[UHH]{I. Institut f\"ur Theoretische Physik, Universit\"at Hamburg, Jungiusstr. 9, 20355 Hamburg, Germany}
	\address[BLN]{Institut f\"ur Physik, Humboldt-Universit\"at zu Berlin, Newtonstr. 15, 12489 Berlin, Germany}
  \cortext[cor]{Corresponding author}

\begin{abstract}
To investigate ferromagnetic semiconductors and insulators, such as the famous EuO, EuS, or CrBr$_3$, we propose a hybridized Kondo--lattice model, where, in addition to the conduction electrons, localized moments (e.g., the $4f$--electrons) are modeled as a strongly correlated band system. The quasi--empty conduction band is weakly filled due to the hybridization term. This activates the  intraatomic exchange coupling between conduction and localized electrons. Temperature--dependent phase diagrams and quasiparticle densities of states are presented for various coupling and hybridization strengths. Moreover, the influence of the one--particle energy of the localized electrons $E_f$ is discussed. A comparison with mean field calculations is given at the end of this work.
\end{abstract}


\maketitle


\section{Introduction}\label{ChapIntro}
Collective magnetism requires the existence of permanent magnetic moments, which arise from either localized or itinerant electrons. The distinction of the magnetic moment's type is of enormous conceptual importance: While itinerant electrons cause band magnetism, particularly known from the classical ferromagnets Fe, Co and Ni, and usually examined within the Hubbard model \cite{Gutzwiller.1963, Hubbard.1963, Kanamori.1963}, local moment magnetism can be found in insulators and semiconductors (such as EuO, EuS and CrBr$_3$ \cite{Ivanov.2007, Kunes.2005, SouzaNeto.2009}), in the huge class of the heavy fermion systems, in diluted magnetic semiconductors (which are promising candidates for spintronic applications \cite{Dietl.2001, Ohno.1999, Rausch.2011, Wolf.2001}), and also in some metallic systems like Gd. Such local moment systems are usually described by spin models (e.g. Heisenberg or Ising model \cite{Ising.1925, Santos.2004}) with coupling constants due to direct exchange or any type of superexchange. It remains an important aspect of modern magnetism fundamental research to find a unified theory which depicts the variety of magnetic phenomena. A small step along this path will be presented in this paper.

Since collective magnetism is exclusively realized in solid state bodies, whose electronic structure is represented by energy bands and gaps that consequently determine the magnetic properties, it seems to be the more natural choice to use band models instead of pure spin models. In the present work, we propose a band model that describes semiconductors and insulators, i.e., local moment systems. For our purposes, the well--known Kondo--lattice model \cite{Kondo.1964}, with local spins replaced by correlated $f$--orbitals, is a formidable starting point as the interaction between subsystems (in our case: localized and itinerant electrons) may lead to magnetic ordering. Note, however, that the interaction is of indirect nature (e.g. Ruderman-Kittel-Kasuya-Yosida interaction \cite{Kasuya.1956, Ruderman.1954, Yosida.1957}, hereafter referred to as RKKY), thus, in search of spin--ordering effects, requiring at least a minimum of electrons in each subsystem. The problem one is now confronted with is that semiconductors, and insulators likewise, have quasi--empty conduction bands, making indirect electronic exchange impossible. In our work, we therefore extend the Kondo--lattice model by a hybridization term, which gives localized electrons the opportunity to virtually transform into conduction band electrons and vice versa. This solves the problem of beforehand absent conduction electrons, hence allowing a virtual RKKY interaction, but also destabilizes the magnetic moments in the different subsystems. As we can read off from our results, the last fact is of great importance for the understanding of the magnetism found with our model proposed.

The setup of this paper is as follows: In Sec. \ref{ChapModel} we present the Hamiltonian, which characterizes the many--body problem that we solve by Green's functions methods in Sec. \ref{ChapTheory}. Within this solution, a set of self--energies is required, but so far undetermined. Approximations to compute the self--energies are presented in Sec. \ref{SpezLsgDerSelfen}, finalizing the theoretical framework. The results and a corresponding discussion are presented in Sec. \ref{ChapResults}. A summary is given in Sec. \ref{ChapSummary}.

\section{Model}\label{ChapModel}
For the above described systems we introduce the following Hamiltonian $H$:
\begin{equation}\label{Hamiltonian}
H = H_s + H_f + H_{f(U)} + H_{sf} + H_V~.
\end{equation}
The kinetic energy of the conduction band electrons is concerned by $H_s$,
\begin{equation}\label{DefHS}
H_s = \Hs~,
\end{equation}
where $\cisd$ ($\cis$) are creation (annihilation) operators of conduction band ($s$-)electrons at lattice site $\boldsymbol{R}_i$ with spin $\sigma$, respectively. The chemical potential is denoted by $\mu$, and $T_{ij}$ are the usual hopping matrix elements, which are related to the Bloch energies $\epsk$ via
\begin{equation}
T_{ij} = \frac1N \sumk \epsk e^{-i\boldsymbol{k}\kl{\boldsymbol{R}_i - \boldsymbol{R}_j}}~.
\end{equation}
For low band occupations, a Coulomb interaction between $s$--electrons can be neglected safely.

Assuming a non-degenerate $f$--level $E_f$, the one--partic\-le energy of the localized $f$--electrons with creation (annihilation) operators $\fisd$ ($\fis$) is represented by
\begin{equation}\label{Hf1}
H_f = \Hfeins~.
\end{equation}
In order to disfavor double occupancies, a large intra\-atomic Coulomb interaction is taken into account by
\begin{equation}\label{Hf2}
H_{f(U)} = \Hfzwei~,
\end{equation}
where $\nfis = \fisd\fis$ is the number operator. Since the $4f$--wave functions' overlap is negligible, a direct $f$--electron exchange is fairly small and hence not part of the Hamiltonian \eqref{Hamiltonian}. Note, moreover, that Eqs. \eqref{Hf1} and \eqref{Hf2} are only valid for systems with a total spin of 1/2, thus being an approximation in modeling the $4f$--levels. This simplification is done in order to keep mathematics on a tractable level, whereas the underlying physical effects are not expected to be affected significantly.

To allow for magnetic ordering, interactions between conduction band and $f$--electrons must be considered. The $s$--$f$--exchange is commonly described by an intraatomic spin--spin--coupling,
\begin{equation}\label{Hsf}
H_{sf} = -J\sumi\boldsymbol{\sigma}_i\cdot\boldsymbol{S}_{i}~,
\end{equation}
where $\boldsymbol{\sigma}_i$ and $\boldsymbol{S}_{i}$ represent the conduction and $f$--electron spin operators at lattice site $\boldsymbol{R}_i$, respectively. The coupling strength is given by $J$, where positive (negative) values of $J$ stand for a preferred (anti)parallel alignment of $s$- and $f$--electron magnetic moments.

The last part of the Hamiltonian in Eq. \eqref{Hamiltonian},
\begin{equation}\label{HamV}
H_V = \HV~,
\end{equation}
allows for (virtual) electronic transitions from the $4f$--level into the conduction band and vice versa, where $V$ is the hybridization strength. While the $s$--$f$--exchange, Eq. \eqref{Hsf}, originates from the non--classical part of the Coulomb interaction between the conduction electrons and the localized $f$--electrons, the hybridization term, Eq. \eqref{HamV}, is a one--particle scattering term which mimics, in the most simple way, the hybridization between the respective bands. While concrete values might be obtained by standard self--consistent band--structure or constraint RPA calculations, there is a priori no reason why those couplings should not co--exist.

As a consequence of the hybridization, the respective average occupation numbers
\begin{equation}\label{DefBes}\begin{split}
n_{s\sigma} = \left\langle \cisd\cis \right\rangle~,&\quad n_{f\sigma} = \left\langle \fisd\fis \right\rangle~,\\
n_s = \sumsig n_{s\sigma}~,&\quad n_f = \sumsig n_{f\sigma}~,
\end{split}\end{equation}
are not necessarily constants when varying a model parameter. Contrary, the total occupation number
\begin{equation}\label{nges}
n = n_s + n_f
\end{equation}
is kept fixed at a constant value by proper adjustment of the chemical potential $\mu$. Since needed for later purposes, we also introduce the dimensionless magnetizations $m_{s,f}$,
\begin{equation}\label{DefMag}\begin{split}
m_s = \sumsig\zs n_{s\sigma}~,\\
m_f = \sumsig\zs n_{f\sigma}~,
\end{split}\end{equation}
using $\zs = \delta_{\sigma\uparrow} - \delta_{\sigma\downarrow}$.

To fix some of the model parameters, we set the on--site energies of the conduction electrons to $T_{ii}=0$ eV. Hence, the conduction band center of gravity defines the energy zero. Energy units are essentially fixed by choosing $W=1$ eV for the width of the conduction band. Furthermore, we assume a strong Hubbard interaction $U_f \gg W$.

The one--particle energy of the $4f$--levels, $E_f$, is a parameter that decisively affects the physics of the model. A stable local magnetic moment on the $f$-levels is formed in the limit $E_f \ll 0$ and $E_f + U_f \gg 0$ where each $f$-level is exactly occupied by one electron. Here, however, we also consider a parameter regime where $E_f$ comes close to the lower edge of the conduction band. This implies the presence of charge fluctuations and thereby the hybridization term, Eq. \eqref{HamV} is activated which eventually generates an effective $f$--$f$ magnetic exchange. Note that this implies that the Schrieffer--Wolff transformation \cite{Schrieffer.1966} does not apply to this parameter regime. Consequently, $H_f + H_{f(U)} + H_V$ cannot be replaced by a local antiferromagnetic exchange which would trivially compete (or cooperate) with $H_{sf}$. 

It is worth mentioning that the model \eqref{Hamiltonian} reduces to the conventional periodic Anderson model if the $s$--$f$--exchange, Eq. \eqref{Hsf}, is neglected ($J=0$). On the other hand, for $V=0$, the model essentially reduces to a fermionized variant of the famous Kondo-lattice model.

\section{Theory}\label{ChapTheory}
In order to determine $n_{s,f}$ and $m_{s,f}$ self--consistently, we attack the many--body problem posed by the Hamiltonian \eqref{Hamiltonian} using Green's function methods. With the definition of the pure one--particle $s$- and $f$--Green's functions,
\begin{equation}\label{GreenDefEM}\begin{split}
\cGFem &= \frac1N\sumij e^{i\boldsymbol{k}\kl{\boldsymbolR_i - \boldsymbolR_j}}\GFkl{\cis}{\cjsd}~,\\
\fGFem &= \frac1N\sumij e^{i\boldsymbol{k}\kl{\boldsymbolR_i - \boldsymbolR_j}}\GFkl{\fis}{\fjsd}~,
\end{split}\end{equation}
as well as with two mixed Green's functions,
\begin{equation}\label{GreenDefEMMIX}\begin{split}
\pGFone &= \frac1N\sumij e^{i\boldsymbol{k}\kl{\boldsymbolR_i - \boldsymbolR_j}}\GFkl{\cis}{\fjsd}~,\\
\pGFtwo &= \frac1N\sumij e^{i\boldsymbol{k}\kl{\boldsymbolR_i - \boldsymbolR_j}}\GFkl{\fis}{\cjsd}~,
\end{split}\end{equation}
all given in $\boldsymbol{k}$--space with $N$ being the number of lattice sites, one easily obtains the following equation of motion:
\begin{equation}\label{EMgl1}\begin{split}
\cGFem =& \frac1E\left(\hbar + \kl{\epsk - \mu}\cGFem + V \pGFtwo +\right.\\
&+ \frac1N\sumij e^{i\boldsymbol{k}\kl{\boldsymbolR_i - \boldsymbolR_j}}\left.\GFkl{\Kom{\cis}{H_{sf}}}{\cjsd}\right)~.
\end{split}\end{equation}
In Eq. \eqref{EMgl1}, the angular brackets represent a higher \linebreak Green's function. To find a solution of the many--body problem, we use the following ansatz according to the rigorous Dyson equation \cite{Nolting.2009}:
\begin{equation}\label{AnsatzSelfenEM}
\Kom{\cis}{H_{sf}} \longrightarrow \slim{m}{}\Sigma^{(s),\text{sf}}_{im\sigma}(E)c_{m\sigma}~,
\end{equation}
which is only valid within the expression for the Green's function $\GFkl{\Kom{\cis}{H_{sf}}}{\cjsd}$. The equation above defines the electronic self--energy $\Sigma^{(s),\text{sf}}_{im\sigma}(E)$. Note, that Eq. \eqref{AnsatzSelfenEM} does not represent an operator identity. Inserting the self--energy ansatz \eqref{AnsatzSelfenEM} into the equation of motion \eqref{EMgl1}, one obtains:
\begin{equation}\label{fastDyson}\begin{split}
\cGFem =& G_{\boldsymbol{k}}^{0}(E) + \frac1\hbar G_{\boldsymbol{k}}^{0}(E)\Sigma^{(s),\text{sf}}_{\boldsymbol{k}\sigma}(E) \cGFem +\\
& + \frac{V}\hbar G_{\boldsymbol{k}}^{0}(E)\pGFtwo \\
=&\cGFemVnull\cdot\kl{1 + \frac{V \pGFtwo}{\hbar}}~,
\end{split}\end{equation}
where $\Sigma^{(s),\text{sf}}_{\boldsymbol{k}\sigma}(E)$ is the Fourier--transformed self--energy, $G_{\boldsymbol{k}}^{0}$ is the correlation--free $s$-Green's function,
\begin{equation}\label{FreeSGF}
G_{\boldsymbol{k}}^{0}(E) = \frac{\hbar}{E + \mu - \epsk}~,
\end{equation}
and where $G^{V=0}_{\boldsymbol{k}\sigma}$ stands for the hybridization--free ($V = 0$) conduction band electrons' Green's function:
\begin{equation}\label{CGFvnull}
\cGFemVnull = \frac{\hbar}{E + \mu - \epsk - \Sigma^{(s),\text{sf}}_{\boldsymbol{k}\sigma}(E)}~.
\end{equation}
Since needed for later purposes, we also introduce the free propagator $G_0(E)$:
\begin{equation}\label{FreeProp}
G_0(E) = \frac1N\sumk G_{\boldsymbol{k}}^{0}(E)~.
\end{equation}

In analogy to the system of conduction band electrons, the equation of motion of the localized electrons can be derived:
\begin{equation}\label{EMgl2}\begin{split}
\fGFem =& \frac1E\left( \hbar + \kl{E_f - \mu}\fGFem + V \pGFone + \right. \\
&+ \frac1N\sumij e^{i\boldsymbol{k}\kl{\boldsymbolR_i - \boldsymbolR_j}}\GFkl{\Kom{\fis}{H_{f(U)}}}{\fjsd} +\\
&+ \left. \frac1N\sumij e^{i\boldsymbol{k}\kl{\boldsymbolR_i - \boldsymbolR_j}}\GFkl{\Kom{\fis}{H_{sf}}}{\fjsd}\right)~.\\
\end{split}\end{equation}
As in Eq. \eqref{AnsatzSelfenEM}, we use the following substitutions for the $f$--electrons:
\begin{equation}\label{KomFAnsatz}\begin{split}
\Kom{\fis}{H_{f(U)}} &\longrightarrow \slim{m}{}\Sigma^{(f),\text{U}}_{im\sigma}(E)\fms~,\\
\Kom{\fis}{H_{sf}} &\longrightarrow \slim{m}{}\Sigma^{(f),\text{sf}}_{im\sigma}(E)\fms~,
\end{split}
\end{equation}
which are only valid within the Green's functions. Defining a hybridization--free $f$--Green's function $F^{V=0}_{\boldsymbol{k}\sigma}$,
\begin{equation}\label{FGFvnull}
\fGFemVnullk = \frac{\hbar}{E + \mu - E_f - \SFUk - \SFk}~,
\end{equation}
with $\Sigma^{(f),\text{U}}_{\boldsymbol{k}\sigma}(E)$ and $\Sigma^{(f),\text{sf}}_{\boldsymbol{k}\sigma}(E)$ being the Fourier--transfor\-med expressions of the self--energies given through Eq. \eqref{KomFAnsatz}, the full $f$--Green's function can be written as
\begin{equation}\label{greenFontheway}
\fGFem = \fGFemVnullk\cdot\kl{1 + \frac{V\pGFone}{\hbar}}~.
\end{equation}

The mixed Green's functions $P^{(1)}_{\boldsymbol{k}\sigma}$ and $P^{(2)}_{\boldsymbol{k}\sigma}$ are yet unknown. Using the substitutions \eqref{AnsatzSelfenEM} and \eqref{KomFAnsatz} in the equations of motion for $P^{(1)}_{\boldsymbol{k}\sigma}$ and $P^{(2)}_{\boldsymbol{k}\sigma}$, and using the hybri\-dization--free Green's functions \eqref{CGFvnull} and \eqref{FGFvnull}, one gets: 
\begin{equation}\label{greenPone}
\pGFone = \frac{V\cGFemVnull}{\hbar}\fGFem~,
\end{equation}
and
\begin{equation}\label{greenPtwo}
\pGFtwo = \frac{V\fGFemVnullk}{\hbar}\cGFem~.
\end{equation}
No higher Green's functions appear in these results, which enables us to solve the system of $s$- and $f$--Green's function equations by inserting Eqs. \eqref{greenPone} and \eqref{greenPtwo} into Eqs. \eqref{fastDyson} and \eqref{greenFontheway}:
\begin{equation}\label{CGFEM}
\cGFem = \left(\frac{1}{\cGFemVnull} - \frac{V^2}{\hbar^2}\fGFemVnullk\right)^{-1}~,
\end{equation}
\begin{equation}\label{FGFEM}
\fGFem = \left(\frac{1}{\fGFemVnullk} - \frac{V^2}{\hbar^2}\cGFemVnull\right)^{-1}~.
\end{equation}

Of course, this represents a reformulation of the full many--body problem only, since expressions for the self--energies or rather the hybridization-free Green's functions $G^{V=0}_{\boldsymbol{k}\sigma}$ and $F^{V=0}_{\boldsymbol{k}\sigma}$ are still missing. We will work them out in the next section.

\section{Self-energies}\label{SpezLsgDerSelfen}
The last sections' preparatory work shifts the focus from the original problem towards the set of (so far undetermined) self--energies, which have to be calculated approximately.

To find an analytical solution, we make an effective medium approach, meaning that the self--energies are \linebreak worked out within the two respective subsystems, assuming $V=0$. The main idea is that hybridization effects are already taken into account for the Green's functions to some extent (see Eqs. \eqref{CGFEM} and \eqref{FGFEM}). Physically speaking, conduction band electrons then encounter an effective $f$--electron potential and vice versa.

For the hybridization--free conduction band system, \linebreak which is described by the Hamiltonian parts \eqref{DefHS} and \eqref{Hsf}, the associated Green's function is given by Eq. \eqref{CGFvnull}. Using the static mean field (MF) approximation, one obtains the following expression for the $s$--electron's self--energy:
\begin{equation}\label{SselfenMFN}
\Sigma^{(s),\text{sf}}_{\boldsymbol{k}\sigma}(E) \equiv \Sigma^{\text{(s),MF}}_\sigma = -\frac{J}{2}\zs\erw{S^z_f}= -\frac{J}4\zs m_f~,
\end{equation}
where $S^z_f$ stands for the $z$--projection of the $f$--system's spin operator $\boldsymbol{S}_f$. In this paper, the mean field result is introduced for comparison only.

A more reliable expression for $\Sigma^{(s),\text{sf}}_{\boldsymbol{k}\sigma}(E)$ was given by Nolting et. al \cite{Nolting.2001}, who found an approximate solution to the self--energy by interpolating between several exactly known cases, such as the zero--bandwidth limit, the ferromagnetically saturated semiconductor and the second--order perturbation theory. The derived self--energy is trustworthy for arbitrary temperatures and coupling strengths $J$, as well as for low band occupations $n_s$. Likewise using ansatz \eqref{AnsatzSelfenEM}, the following self--energy was obtained:
\begin{equation}\label{isaformfn}\begin{split}
\Sigma^{(s),\text{sf}}_{\boldsymbol{k}\sigma}(E) =& - \frac{J}{4} \zs m_f ~ + \\
& + \frac{J^2}{4}\cdot\frac{S_f(S_f + 1) - \frac{\zs m_f}{2}\kl{\frac{\zs m_f}{2} + 1}}{\left(G_0\kl{E - \frac{J}4 m_f \zs}\right)^{-1} - \frac{J}{2}}~.
\end{split}\end{equation}
Here, $G_0$ is given by Eq. \eqref{FreeProp} and $S_f$ is the $f$--spin quantum number. As mentioned in Sec. \ref{ChapModel}, we treat the $f$--band as a spin--1/2--system, i.e. $S_f=1/2$. Note, however, that Eq. \eqref{isaformfn} is not restricted to this case.

The next step is finding the self--energies $\Sigma^{(f),\text{U}}_{\boldsymbol{k}\sigma}(E)$ and $\Sigma^{(f),\text{sf}}_{\boldsymbol{k}\sigma}(E)$ of the hybridization--free local moment system. Eq. \eqref{FGFvnull} shows, that knowing the $f$--electron's Green's function is sufficient for obtaining the sum of the wanted self--energies. The latter are approximated by the self--energies of the fermionized Kondo--lattice model in the zero--bandwidth limit. The resulting self--energy is $\boldsymbol{k}$--\linebreak independent. This approximation is exact for the case of an empty conduction band and should remain reliable for low conduction band filling.

Starting once again with a mean field (MF) treatment of the local moment system, one obtains a ($\boldsymbol{k}$--inde--pendent) $f$--electron's Green's function which is equivalent to the solution of the zero--bandwidth Hubbard model \cite{Nolting.2009}:
\begin{equation}\begin{split}
\fGFmfnN =& \hbar\left(\frac{1 - \nfms}{E + \mu - E_f + \frac{J}{4}\zs m_s}\right. + \\
& + \left.\frac{\nfms}{E + \mu - E_f + \frac{J}{4}\zs m_s - U_f}\right)~.
\end{split}\end{equation}
As mentioned before, this result will be primarily important for later comparison with the effective medium approach. The theory of the latter, of course, also requires a local moment Green's function, on which we will focus in the following paragraphs.

The system turns out to be exactly solvable in the zero--bandwidth limit \cite{Nolting.1984}. This yields the following Green's function:
\begin{equation}\label{Fvnull}
\fGFemVnullk \equiv \fGFemVnull = \slim{j=1}{4}\frac{\alpha_{j\sigma}}{E - E_{j}}~.
\end{equation}
Coulomb- and $s$--$f$--interaction are responsible for a splitting into four sublevels, which are located at energy positions $E_{j=1,\ldots,4}$,
\begin{equation}\label{fpos}\begin{split}
E_1 =& T_0 - \mu - \frac{J}2 \cdot S_c~,\\
E_2 =& T_0 - \mu + \frac{J}2 \cdot (S_c + 1)~,\\
E_3 =& T_0 - \mu + U - \frac{J}2 \cdot (S_c + 1)~,\\
E_4 =& T_0 - \mu + U + \frac{J}2 \cdot S_c~,
\end{split}\end{equation}
each of them connected to a spectral weight $\alpha_{j\sigma}$, where $j=1,\ldots,4$:
\begin{equation}\begin{split}
\alpha_{1\sigma} =& \frac1{2S_c + 1}\kl{S_c + 1 + \frac{\zs m_s}2 + \DMS - (S_c + 1)\nfms}~,\\
\alpha_{2\sigma} =& \frac1{2S_c + 1}\kl{S_c - \frac{\zs m_s}2 - \DMS - S_c\nfms}~,\\
\alpha_{3\sigma} =& \frac1{2S_c + 1}\kl{S_c\nfms - \DMS}~,\\
\alpha_{4\sigma} =& \frac1{2S_c + 1}\kl{\DMS + (S_c + 1)\nfms}~.
\end{split}\end{equation}
In the equations above, $S_c=1/2$ stands for the total spin of the conduction band system. The average occupation numbers $\nfms$ and conduction band magnetization $m_s$ were already defined in Eqs. \eqref{DefBes} and \eqref{DefMag}, whereas the parameter $\Delta_\sigma$ is a higher correlation function:
\begin{equation}
\Delta_\sigma = \left\langle S_c^\sigma \fmsd \fs\right\rangle + \zs\left\langle S_c^z \ns \right\rangle~.
\end{equation}
Here, $S_c^\sigma = S_c^x + i\zs S_c^y$ is the ladder operator for a conduction electron's spin $\boldsymbol{S}_c$, and $S_c^z$ its $z$--component. In the conventional Kondo--lattice model, this two--particle correlation can, surprisingly, be expressed via the single $f$--electron's Green's function \cite{Nolting.2003}:
\begin{equation}\label{DefDS}
\Delta_\sigma = \frac1{J\pi N}\slim{\boldsymbol{k}}{}\ilim{-\infty}{\infty}f_-(E)(E-E_f)\text{ Im } F_{\boldsymbol{k}\sigma}(E - \mu) dE~,
\end{equation}
where $f_-(E)$ is the Fermi function. By combining Eq. \eqref{Fvnull} with Eq. \eqref{DefDS}, and after tedious, but straightforward work, one obtains $\DS$, as a functional of $\nfs$, $\ncs$ (and thus $m_s$) and $\DS$ itself. Instead of calculating $n_{s,f\sigma}$ within the hybridization--free subsystems, we chose to evaluate them in the full system in order to allow for a better inclusion of hybridization effects. We therefore use the respective Green's functions \eqref{CGFEM} and \eqref{FGFEM}, as well as the spectral theorem \cite{Nolting.2009}:
\begin{equation}\label{SpectralTheorem}\begin{split}
\ncs =& -\frac1\hbar\ilim{-\infty}{\infty}f_-(E)\frac1{N\pi}\sumk \text{ Im } G_{\boldsymbol{k}\sigma}(E - \mu)dE~,\\
\nfs =& -\frac1\hbar\ilim{-\infty}{\infty}f_-(E)\frac1{N\pi}\sumk \text{ Im } F_{\sigma}(E - \mu)dE~.
\end{split}\end{equation}
The $\boldsymbol{k}$--summation in Eq. \eqref{SpectralTheorem} can be replaced by an integration over $x$, weighted by the free Bloch density of states $\varrho_0$:
\begin{equation}\label{freeBloch}
\rN = \frac1N\sumk\delta\kl{x - \epsk}~.
\end{equation}

In self--consistent, numerical calculations, the occupation numbers $n_{s,f\sigma}$ (and hence the magnetizations, see Eq. \eqref{DefMag}) can now be evaluated. A non--vanishing total magnetization,
\begin{equation}
m_{\text{tot}} = m_s + m_f~,
\end{equation}
indicates a ferromagnetic phase.

\section{Results and Discussion}\label{ChapResults}
Referring to systems like EuX (X=O,S,Se,Te), a face-centered cubic (f.c.c.) lattice structure is considered. Calculations are done for an intraatomic Coulomb interaction among the localized electrons of $U_f = 10$ eV according to experimental data (see e.g. \cite{Nolting.1988}). Recall that the conduction band center of gravity is fixed at $T_0=0$ eV and that the width of the conduction band is assumed as $W= 1$ eV. Keep also in mind, that all calculations are performed for a total spin of $1/2$ for both conduction band ($S_c$) and local moment system ($S_f$).

We divide this Section into several parts, beginning with metallic systems without hybridization influences\linebreak (part \ref{SecMR}). Part \ref{SecKLL} addresses the case of a semiconductor or insulator but still $V=0$. Finally, the results of the full theory with finite hybridization are discussed and also compared with mean field calculations in part \ref{SecGenCase}.

\subsection{Magnetic response of a metallic conduction band for V=0}\label{SecMR}
The first special case to be discussed is the metallic Kondo--lattice limit \cite{Kondo.1964, Nolting.1999}, where the hybridization is switched off ($V=0$). The conduction band is taken as half--filled.

\begin{figure}[!ht]
 \centering
	\subfigure[$J=0.1$ eV]{\label{fig:FakeJ01}\includegraphics[width = 0.48\textwidth]{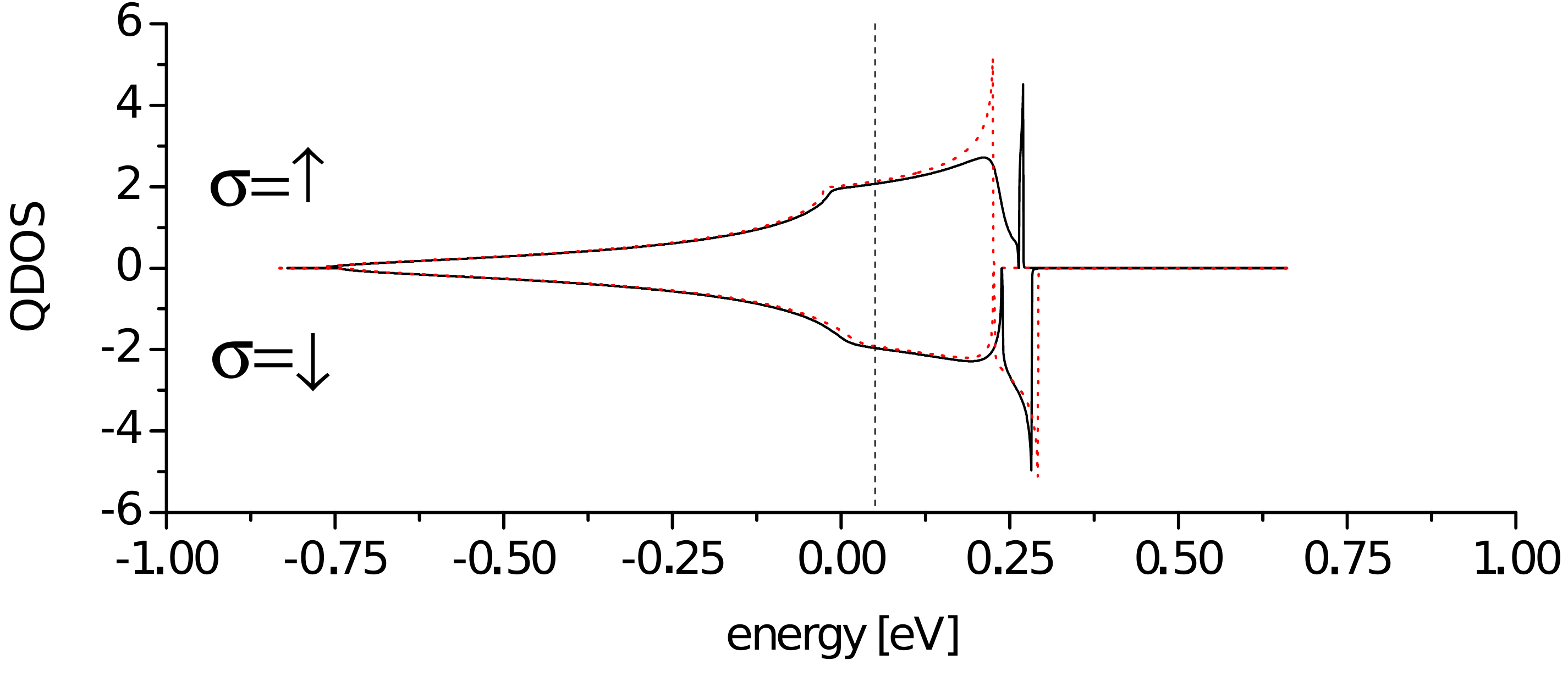}}
	\subfigure[$J=1$ eV]{\label{fig:FakeJ1}\includegraphics[width = 0.48\textwidth]{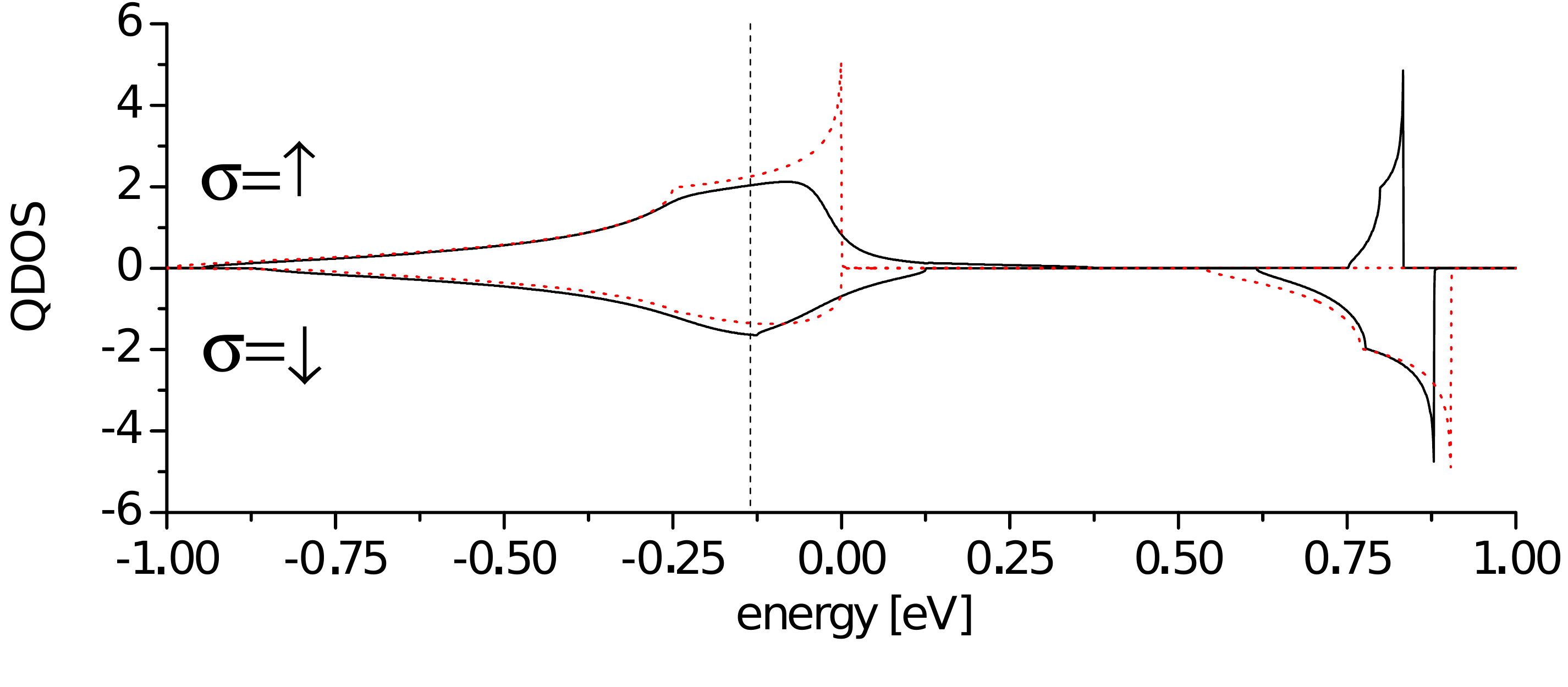}}
\caption{Spin--resolved quasiparticle densities of states (QDOS) of the conduction electrons in the Kondo--lattice model for a half-filled conduction band and for different coupling strengths $J$ with fixed $f$--magnetizations $m_f=0.5$ (solid black lines) and $m_f=1$ (dotted red lines). The chemical potential $\mu$ is marked by the dashed vertical line. $\mu$ is only very slightly temperature--dependent, i.e. dependent on $m_f$.}
\label{fig:FakeMf}
\end{figure}

Some characteristic features of the half--filled Kondo--lattice model can be understood by looking at the quasiparticle density of states shown in Fig. \ref{fig:FakeMf}. For simplicity, we assume to have a fixed $f$--electron's magnetization $m_f$, which - instead of being calculated self--consistently - is treated here as an external parameter. Its temperature dependence is given via a Brillouin function with a Curie temperature of 300 K.

The spin resolved quasiparticle densities of states \linebreak shown in Fig. \ref{fig:FakeMf} split into two subbands which are roughly located at energies $-JS_f/2$ and $J(S_f + 1)/2$. This is due to the interpolating self--energy ansatz Eq. \eqref{isaformfn} and is reminiscent of the splitting found in the zero--bandwidth limit. A known shortcoming of the self--energy Eq. \eqref{isaformfn} is that even for weak coupling strengths $J$ the gap between both subbands persists (see Fig. \ref{fig:FakeJ01}). Nevertheless, this does not seriously affect our weak--$J$ calculations since the gap is negligibly narrow.

By comparing the results for different $m_f$, one finds the expected proportionality to the conduction band's polarization $m_s$. Note, that decreasing $m_f$ goes along with an enhancement of temperature. Moreover, the spectral weight of the upper $\uparrow$--band decreases as the temperature is reduced, vanishing in the case of $T=0$ K. This effect is similar to the special case of the ferromagnetically saturated semiconductor (for further details see Refs. \cite{Henning.2012, Nolting.2001}). In such saturated systems ($m_f=1$), the $\uparrow$--electron cannot flip its spin, and thus the lower $\uparrow$--quasiparticle density of states has the shape of the free (f.c.c.)--Bloch density of states. Of course, this is only valid for the saturated electron sort.

The quasiparticle density of states changes qualitati\-vely in the weak coupling regime, since both subbands (almost) merge (see Fig. \ref{fig:FakeJ01}). When increasing the coupling strength $J$, a correlation gap occurs, which separates the upper band from the Fermi edge (see Fig. \ref{fig:FakeJ1}). Then, exclusively the lower band determines the electronic properties, which results in a rather constant magnetization $m_s$. Later we will see, that this saturation effect plays an important role in our model.

These observations have been made for a given $f$--mag--\linebreak netization and illustrate the reaction to an effective external magnetic field. Of course, also the back reaction of the $s$--magnetization on the local moment system must be considered in addition to get a self--consistent picture.

\subsection{Self--consistent treatment of the Kondo--lattice limit (V=0)}\label{SecKLL}
Let us now come back to the full theory for $V = 0$. Here, $m_f$ is no longer treated as an external parameter. Both $m_f$ and $m_s$ are rather obtained from the subsystem occupation numbers which are calculated self--consistently.
To address the case of a semiconductor or insulator, the total occupation number is fixed at $n=1$ by adjusting the chemical potential accordingly. Let us discuss two parameter regimes:

If $E_f \ll \mu \ll E_f + U$, there are well-defined local magnetic moments formed in the $f$-electron system. This, however, implies that $n_f = 1$ and that the conduction band is empty. A ferromagnetic state cannot emerge in this case.

Ferromagnetism is possible for $V=0$ and $n=1$, in principle, if there is a non-vanishing energy overlap of the lower $f$-electron subband with the (split) conduction bands. Otherwise, conduction band electrons are missing to mediate an effective magnetic exchange between the local moments in the $f$--subsystem. As the total filling is fixed at $n=1$, the total magnetization $m_s + m_f \le 1$. On the other hand, an overlap between $s$- and $f$--type bands not only implies $n_s >0$ but at the same time $n_f <1$. This means that there are charge fluctuations which destabilize the formation of $f$--magnetic moments and therefore disfavor ferromagnetic order.
Concluding, we expect that ferromagnetic order is unlikely for $V=0$ (and $n=1$), irrespective of the $f$--level position $E_f$. 

To address this question numerically, we have studied the model in an $E_f$--range where ferromagnetism could in principle be possible according to the discussion above. We have scanned the entire regime of coupling strength $J$. If at all, ferromagnetism was expected for strong $J$. However, only paramagnetism has been found, irrespective of the value of the coupling strength $J$. From the numerical point of view, this result crucially depends on the fact that, unlike in similar approaches (e.g. \cite{Henning.2009}), our local moment system is treated fully self--consistently.

We conclude that the self--consistent approach well describes the essential spin--flip processes preventing a ferromagnetic order of local moment and conduction band electrons.

\subsection{General case ($V > 0$)}
\label{SecGenCase}
As the discussion above has shown, a hybridization--free system cannot explain the ferromagnetic nature of magnetic semiconductors and insulators, since there are no charge carriers within the conduction band (due to the finite gap between the latter and the local moment energy positions). By switching on the hybridization--term, hybridization--induced virtual transitions of localized $f$--electrons into the conduction band lead to a non--vanishing amount of charge carriers that participate in $s$--$f$--interaction processes. These are responsible for the alignment of the electron spins.

The numerical studies presented below are based on an implementation of the full theory, as detailed in Secs. \ref{ChapTheory} and \ref{SpezLsgDerSelfen}. We again fix the total occupation number at $n=1$, but systematically vary the energy position of the $f$--level $E_f$ as well as the coupling strength $J$ and the hybridization $V$. For the Hubbard--$U$, we choose $U=10$ eV as above.

\begin{figure}[!ht]
 \centering
 	\subfigure[$J=0.2$ eV]{\label{fig:QDOSV03J02}\includegraphics[width = 0.48\textwidth]{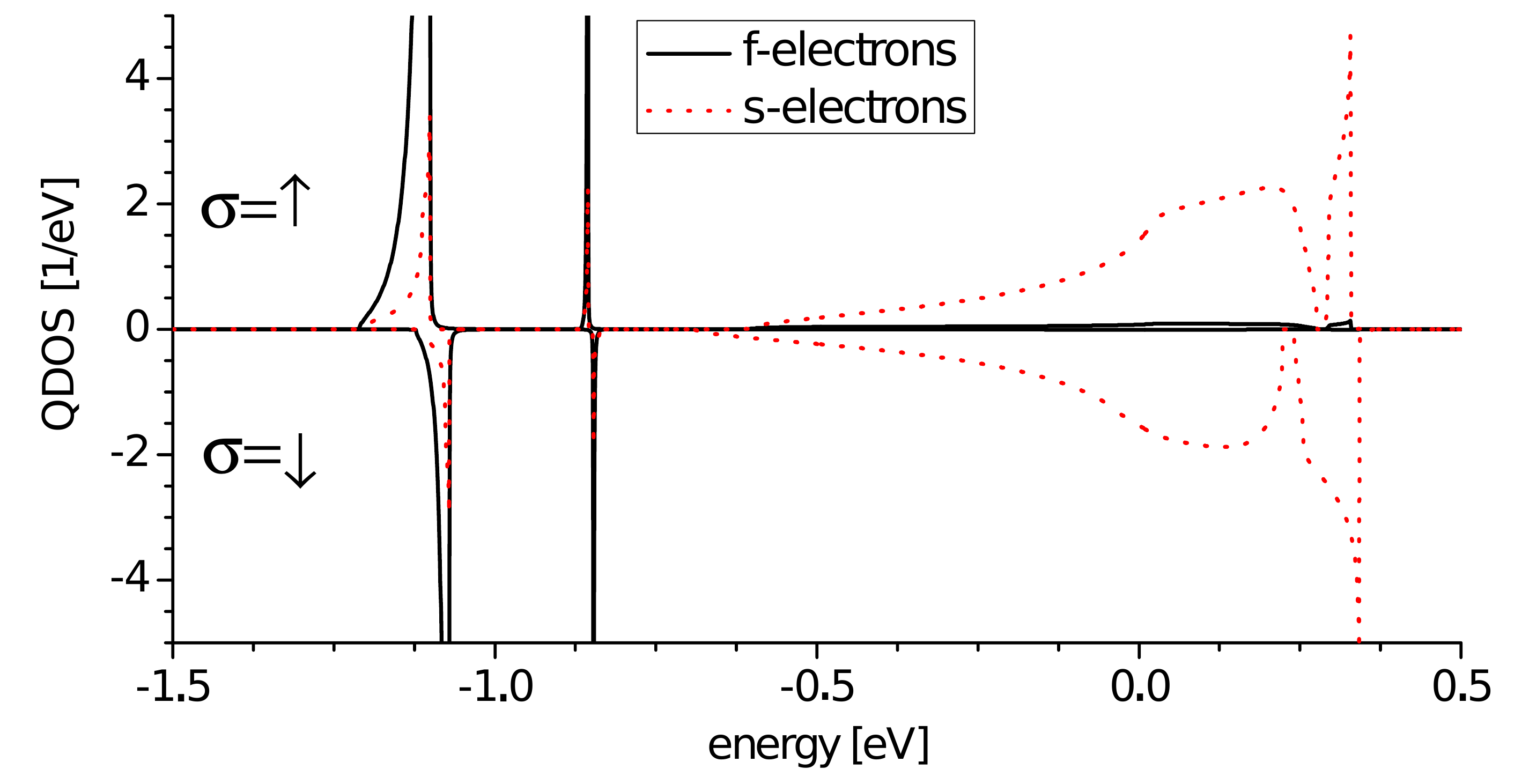}}
	\subfigure[$J=1$ eV]{\label{fig:QDOSV03J1}\includegraphics[width = 0.48\textwidth]{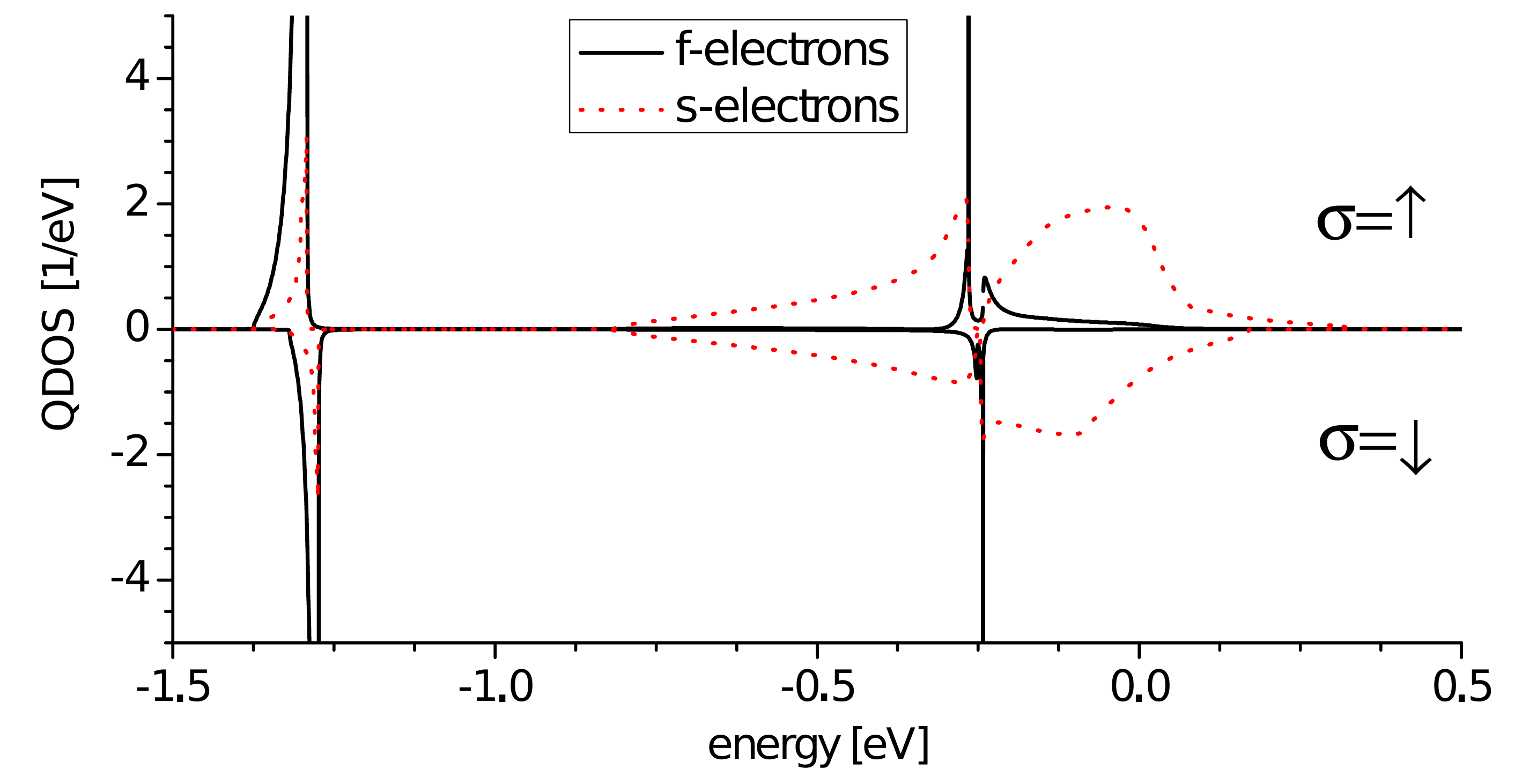}}\\
\caption{Quasiparticle densities of states (QDOS) in the full system for different coupling strengths $J$ with fixed hybridization $V=0.3$ eV. The local moment energy lies at $E_f=-1$ eV. Coulomb repulsion: $U=10$ eV. Note that $f$--levels located at $\sim E_f + U$ are not shown. The chemical potential $\mu$ is located at $\mu=-1.074$ eV for $J=0.2$ eV (upper figure) and $\mu=-1.276$ eV for $J=1$ eV (lower figure).}
\label{fig:QDOS}
\end{figure}

A typical example for the resulting quasiparticle densities of states is shown in Fig. \ref{fig:QDOS}. For two different values of $J$, referring to the weak- and the intermediate coupling regime, a ferromagnetic solution is easily stabilized. This is achieved with a small but finite hybridization $V=0.3$ eV. Note, that with $E_f=-1$ eV there would not be any energy overlap of $s$- and $f$-subsystem bands for $V=0$. The fact, that the $s$- and $f$--quasiparticle densities of states have the same support is a mere consequence of the finite hybridization.

\begin{figure}[ht!]
 \centering
 	\subfigure[$m_{\text{tot}}$ for fixed $s$--$f$--interaction strength $J=1$ eV]{\label{fig:MtotV}\includegraphics[width = 0.48\textwidth]{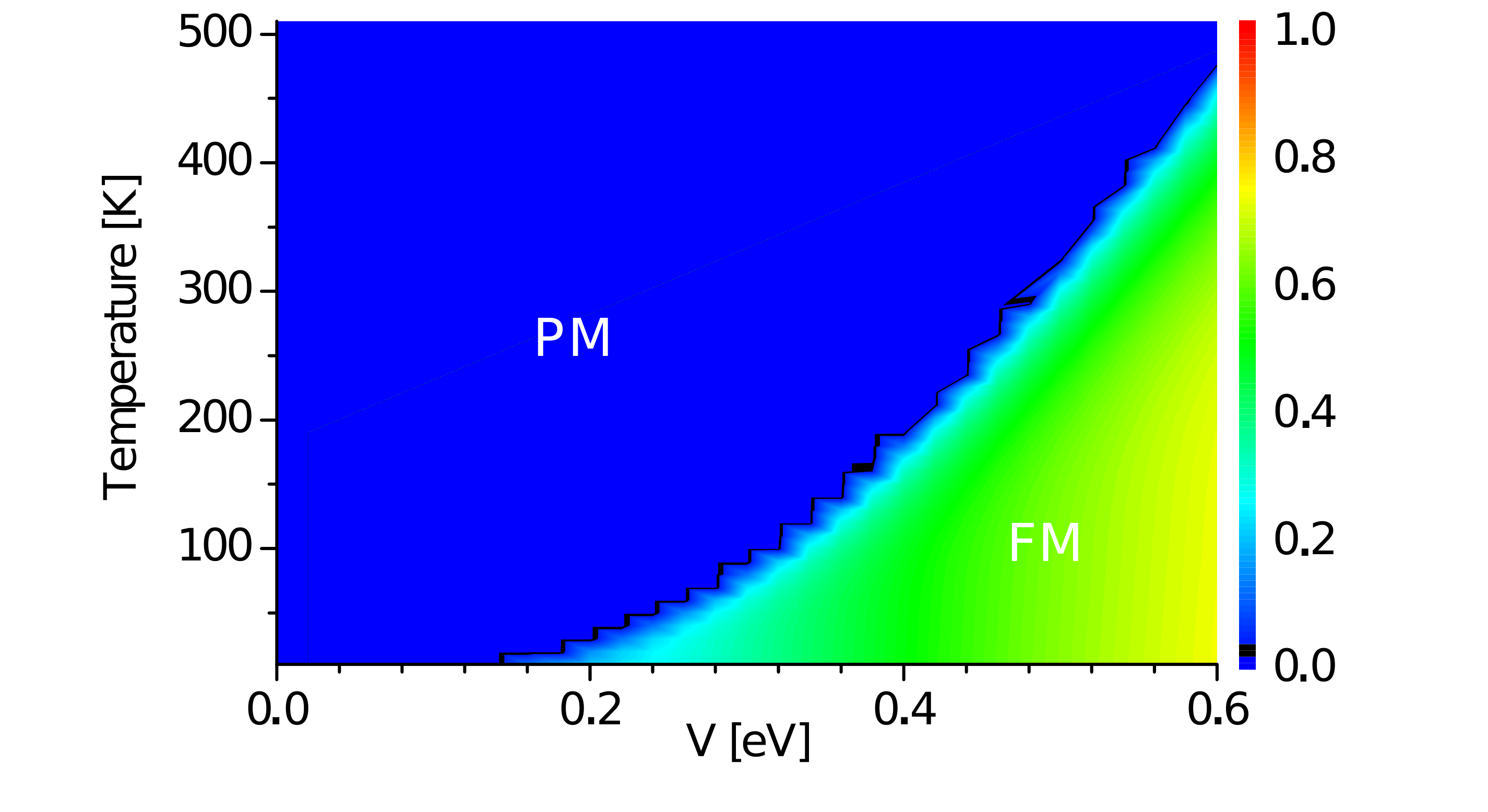}}
	\subfigure[$m_{\text{tot}}$ for fixed hybridization $V=0.3$ eV]{\label{fig:MtotJ}\includegraphics[width = 0.48\textwidth]{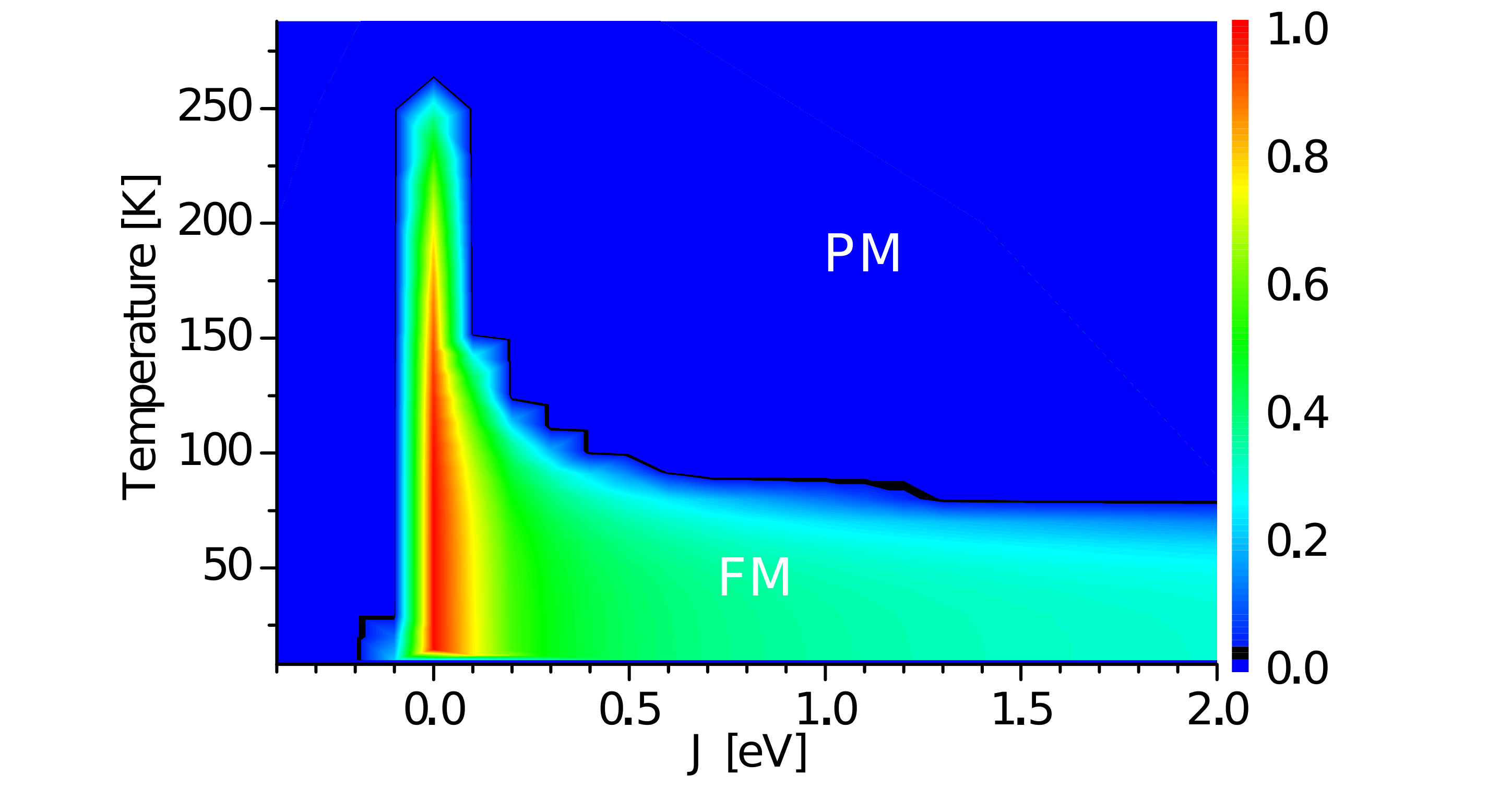}}\\
\caption{Total magnetization $m_{\text{tot}}= m_s + m_f$, represented by the color scale, for various $s$--$f$--interaction strengths $J$ and hybridizations $V$. $E_f$ is fixed at -1 eV and the total occupation number is fixed at $n=1$. FM is the ferro-, PM the paramagnetic phase, both separated by the black line, which marks the Curie temperature.}
\label{fig:MagnetizationTotalSystem}
\end{figure}

Variation of the $s$--$f$--coupling strength $J$ and the hybridization $V$ has significant influences on the total magnetization, and thus on the Curie temperature, as can be seen in our results in Fig. \ref{fig:MagnetizationTotalSystem}. Summarizing, our main results are:
\begin{enumerate}
	\item[(i)] Ferromagnetism occurs for finite $V$ only.
	\item[(ii)] The Curie temperature increases with hybridization strength $V$.
	\item[(iii)] Around $J=0$ eV the Curie temperature takes its highest values and the system is highly polarized in the low--temperature regime.
	\item[(iv)] For large $J$, the Curie temperature saturates.
	\item[(v)] Positive values of $J$ show ferromagnetism, negative ones commonly paramagnetism.
\end{enumerate}

The limiting case $V=0$ eV is a clearly paramagnetic situation, as discussed in Sec. \ref{SecKLL}, see Fig. \ref{fig:MtotV}. For $V>0$ eV, virtual $s$- and $f$--electron transitions occur, causing a broadening of the quasiparticle densities of states, which is clear since the hybridization--term in the Hamiltonian \eqref{HamV} resembles a hopping process. Moreover, the spin--resolved quasiparticle densities of states are repulsively shifted (as $V$ acts like a perturbation), see Fig. \ref{fig:QDOS}. The latter effect increases the spin asymmetry in the quasiparticle densities of states, and favors ferromagnetism. This explains point (ii) of the above list.

Regarding the $J$--dependence, let us emphasize once more that conventional coupling mechanisms such as \linebreak RKKY in the weak--$J$ regime or double exchange in the strong--$J$ regime do not apply here. The point is that the total occupation number $n$ is fixed to 1. Therefore, any increase of the $f$--electron number implies a decrease of the conduction electron number. This is found to destabilize the magnetic order.

For $J=0$ eV, ferromagnetism is found to be very stable with a high Curie temperature (see Fig. \ref{fig:MtotJ}). Ferromagnetic order originates in this case from the strongly correlated $f$--band where the Hubbard--$U$ is much larger than the effective bandwidth resulting from the virtual hybridization processes introduced by the finite $V$.

A finite $J$ brings up two contrary effects, a splitting of $\uparrow$- and $\downarrow$--quasiparticle densities of states due to the linear--in--J (Ising-)term on the one hand and a reduction of ordered spin structures owing to spin--flip processes given by terms with quadratic $J$ on the other hand. From our results we must conclude, that the latter dominates. As spin--flip processes close the gap between $\uparrow$- and $\downarrow$--quasiparticle densities of states, the system's total polarization decreases. Nevertheless, the Curie temperature saturates for $J\rightarrow \infty$ and a finite but small magnetic moment is retained (see point (iv)).

Last but not least, the sign of $J$ is quite important when looking for ferromagnetic phases. Starting off with the hybridization--induced shifts of the $\uparrow$- and $\downarrow$--quasi--particle densities of states, negative coupling strengths produce an Ising term that splits the densities of states into the opposite direction, hence reducing the spin asymmetry. As the sign of $J$ is of importance only in the Ising term, the paramagnetic phase for $J<0$ and the ferromagnetic phase for $J>0$ result from this first--order--in--$J$ term only, just like static mean field calculations would predict.

Another way of understanding our results is as follows: For a system with non--vanishing hybridization and a majority of $f_\uparrow$--electrons, a majority of occupied $s_\uparrow$--states evolve due to hybridization. Such a parallel alignment of most electrons is either energetically favorable ($J>0$ eV) or unfavorable ($J<0$ eV). In the latter case, both electron sorts tend to order antiparallelly, which makes the occupation of states with opposite spin configuration more attractive. The less effective level--repulsion effect described above leads to a decrease of the spin asymmetry.

\begin{figure}[!ht]
 \centering
 	\subfigure[Fixed parameters $V=0.3$ eV and $n=1$.]{\label{fig:TcEfV03}\includegraphics[width = 0.48\textwidth]{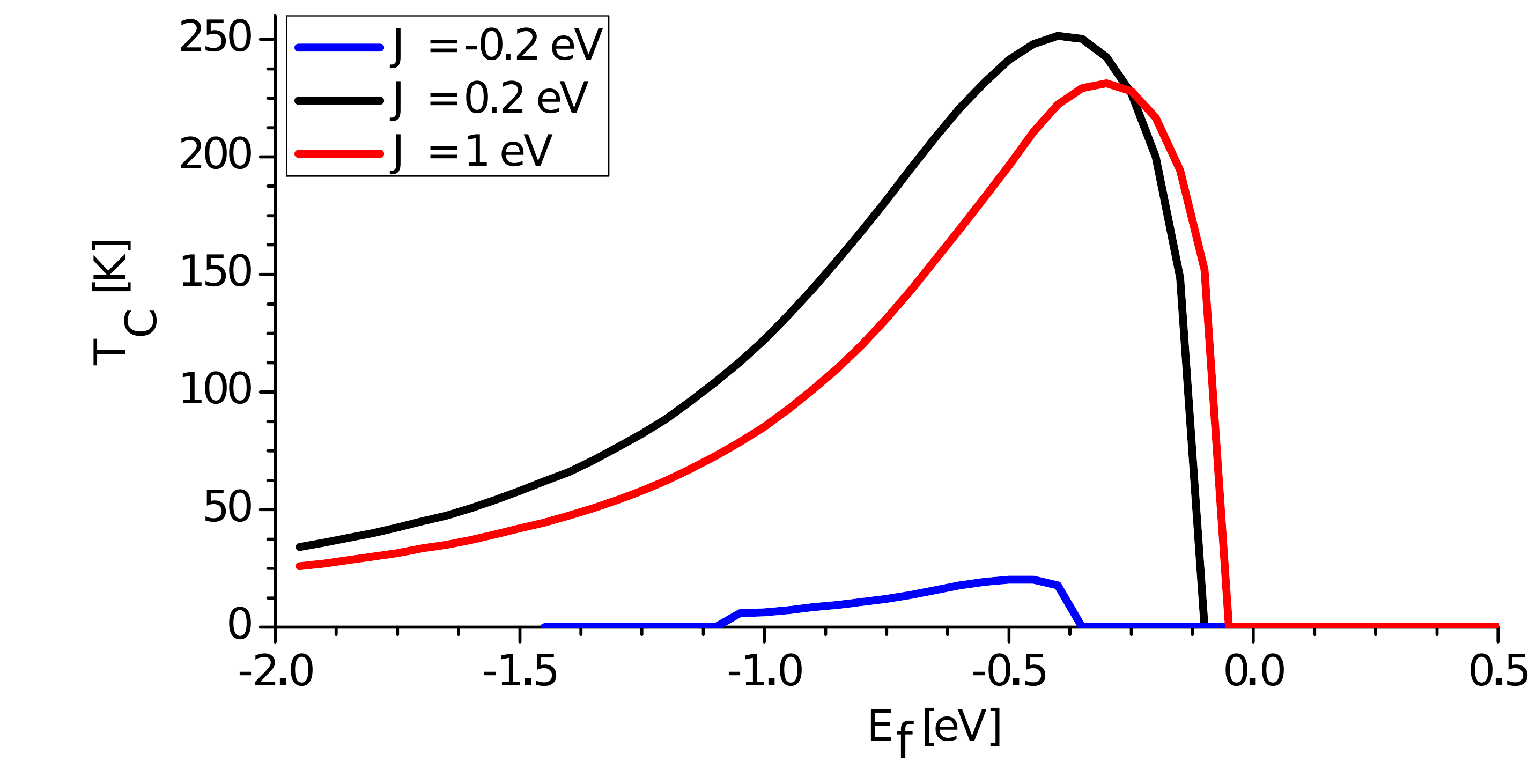}}
	\subfigure[Fixed coupling $J=0.2$ eV and hybridization $V=0.3$ eV.]{\label{fig:TcEfN}\includegraphics[width = 0.48\textwidth]{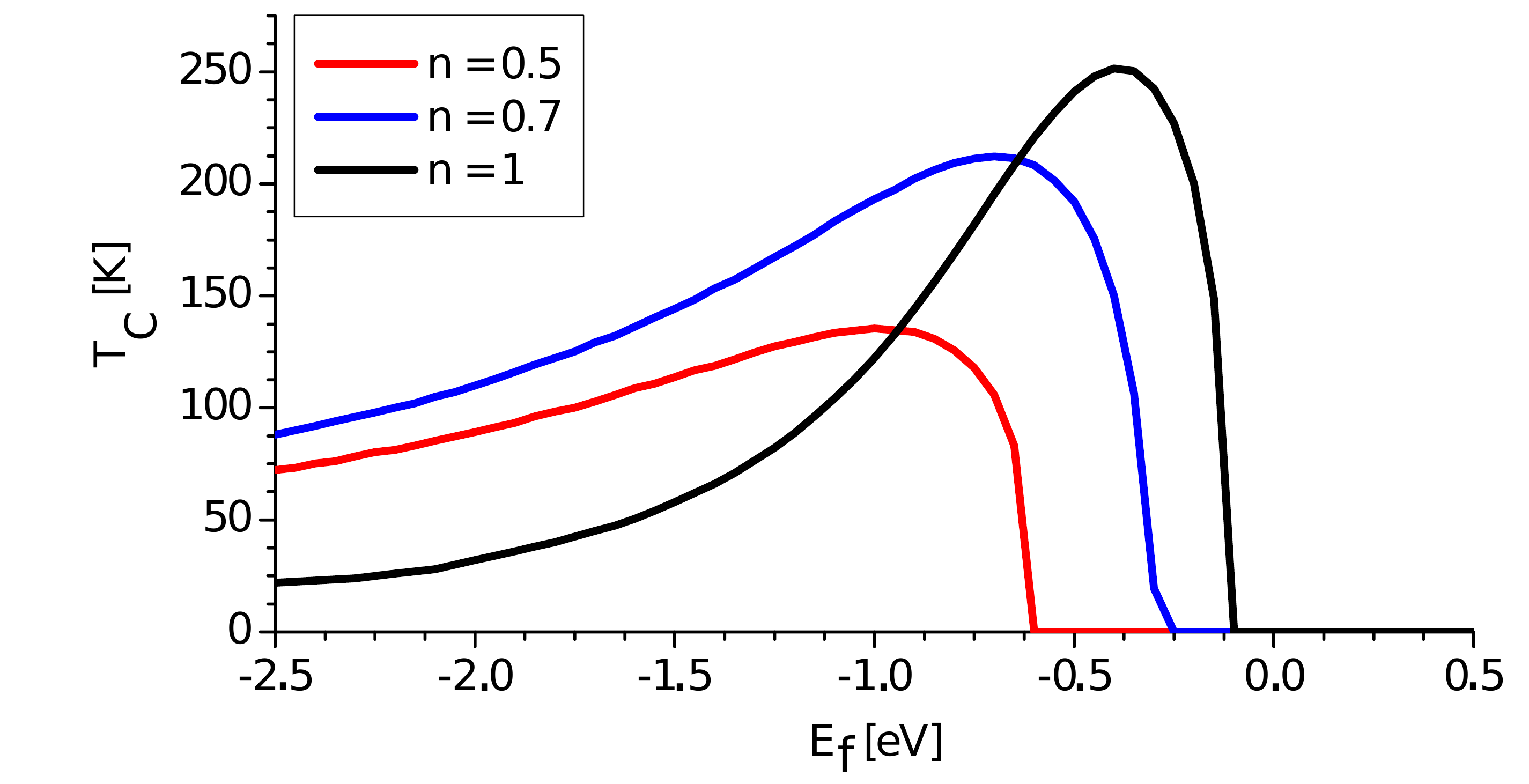}}
\caption{Curie temperature as a function of the local moment energy level position $E_f$ for different parameters such as the $s$--$f$--coupling $J$ (Fig. \ref{fig:TcEfV03}) and the occupation number $n$, Fig. \ref{fig:TcEfN}.}
\label{fig:TcEf}
\end{figure}

Ferromagnetism in this system strongly depends on $E_f$, i.e. on the energy distance between the local moment and conduction band system. Closing (opening) the gap between both subsystem's densities of states increases (decreases) hybridization influences. Hence, shifting $E_f$ leads to an effective hybridization, which eventually changes the magnetization and the Curie temperature (see Fig. \ref{fig:TcEfV03}). With increasing $E_f$ the $s$- and $f$--subbands start to overlap. On the one hand, this produces charge carriers in the conduction band which promote ferromagnetic order as they mediate an effective magnetic coupling between the $f$--electrons' spins. On the other hand, increasing $E_f$ simply introduces charge fluctuations in the $f$--levels and destroys the local $f$--moments. This leads to a rapid breakdown of the ferromagnetic phase as soon as the $f$--level is pushed into the conduction band. A similar $E_f$ dependence, though with different values of the Curie temperature, has been found in previous works using the mean field ansatz (see particularly Refs. \cite{Matlak.1984, Nolting.1987}).

As our model is not necessarily restricted to semiconductors and insulators (such as the Europium chalcogenides), we additionally investigate the physics for different occupation numbers $n$. As is seen in Fig. \ref{fig:TcEfN}, a spectacular influence on the Curie temperature is found. Again, one needs to distinguish between the open gap situation ($E_f$ below the conduction band) and the closed gap regime ($E_f$ inside the conduction band). In the first case, decreasing $n$ means to introduce more and more charge carriers in the conduction band system that mediate an effective coupling of the $f$--electron moments. Hence, $T_C$ increases, see Fig. \ref{fig:TcEfN}. Further decrease of $n$ reduces the number of occupied $f_\uparrow$--states, and will finally lower the Curie temperature.

When pushing the $f$--level into the conduction band (second case), the ferromagnetic phase breaks down faster for small occupation numbers than for larger ones. This is due to the filling order, starting now with the conduction band (which in this situation gives the lowest states to be occupied). As soon as all $s$--electron states are filled, the remaining electrons will occupy the $f$--states given at energy $E_f$ within the conduction band. The smaller $n$ is chosen, the fewer $f$--states will be occupied, hence reducing the corresponding magnetization $m_f$.

\begin{figure}[!htbp]
	\centering
		\includegraphics[width=.48\textwidth]{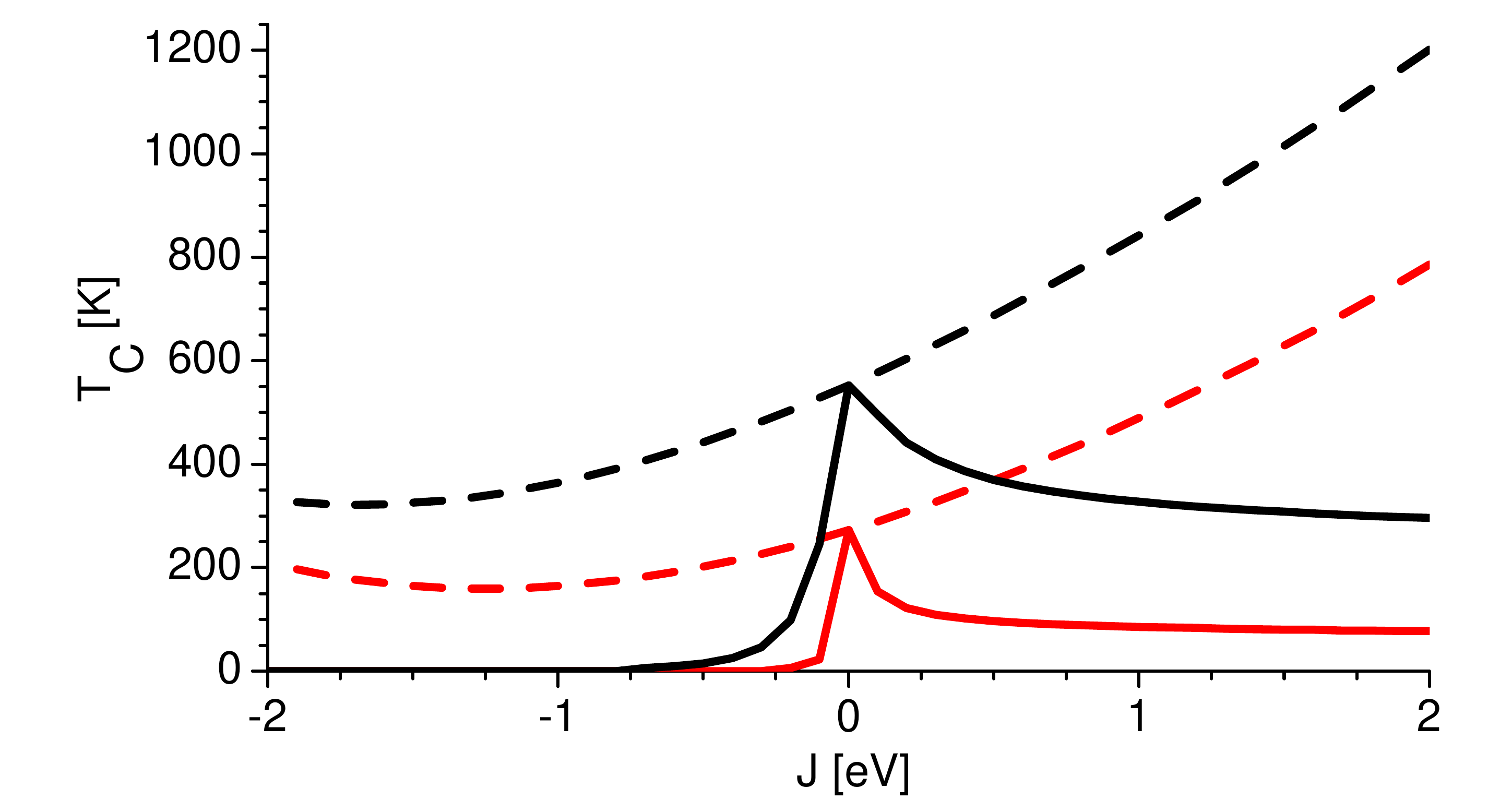}
	\caption{Curie temperature as a function of the coupling strength $J$ for mean field (dashed lines) and effective medium (solid lines) approach. Red graphs are calculated for constant hybridization $V=0.3$ eV, black lines for $V=0.5$ eV. $E_f$ is fixed at -1 eV.}
	\label{fig:MF}
\end{figure}

As a last result, we want to point out the differences between mean field theory (see also Refs. \cite{Matlak.1984, Nolting.1987}) and our effective medium approach. As seen in Sec. \ref{SpezLsgDerSelfen}, our ansatz goes beyond mean field as we allow for higher--order contributions of the $s$--$f$--electronic correlation. Fig. \ref{fig:MF} shows that both theories give the same results in case of $J=0$ eV. As the self--energy vanishes in both cases, this is a trivial result (see Eqs. \eqref{SselfenMFN} and \eqref{isaformfn}). Strong deviations are found for $J\neq0$ eV. All differences are due to the quadratic--in--J term contributions. The saturation effect for large $J$, for instance, is not found in the mean field results. The Curie temperature rather increases with increasing $J$ without any limit.

Even with the mean field treatment of the spin--spin coupling, Eq. \eqref{Hsf}, and even at $J=0$, there is a non-trivial correlation effect that is induced by $U_f$. The ferromagnetic order in the case $J=0$ solely originates from this Hubbard interaction. In fact, due to the charge fluctuations introduced by $V$, the $f$--level develops a finite but small width such that one can effectively think of a strongly correlated Hubbard model with a small band width which apparently sustains a stable ferromagnetic phase.

\section{Summary}\label{ChapSummary}
We have presented a band model for describing ferromagnetic semiconductors and insulators. Referring to local moment systems like EuO, EuS, Gd, etc., the model consists of two subsystems, namely a band of conduction ($s$-)electrons on the one, and strongly localized $f$--electron levels on the other hand. With respect to the above named class of substances, local moment ferromagnetism requires singly occupied $f$--levels, which can be achieved by use of sufficiently large on--site Coulomb interaction. Although the conduction band is rather sparsely occupied, or even empty in the extreme case of temperatures close to absolute zero, we expect it to be necessary for ferromagnetism in local moment systems. We therefore include a hybridization term which allows for (virtual) electron transformations between both subsystems. Last but not least, the two subsystems are given the chance to interact with each other ($s$--$f$--interaction). A hybridized Kondo--lattice model perfectly matches the requirements above and is, hence, used in our work.

We solve the corresponding many--body problem by firstly treating both subsystems separately under exclusion of hybridization, and then using the latter to connect the two systems. Since each subsystem then resembles an effective influence on its counterpart, we call this an effective medium approach. In order to solve the system of conduction band electrons, a self--energy containing the $s$--$f$--interaction influences is required. We choose an (approximate) expression for the self--energy, taken from \cite{Nolting.2001b}, which results from an interpolation between a large amount of exactly solvable special cases, such as second order perturbation theory, the zero--bandwidth limit and the magnetic polaron. As for the local moment system, the strong localization of $f$--electron's wave functions allows us to use an analytically exact zero--bandwidth solution. By connecting the subsystems with finite hybridization in our full theory, we are able to calculate occupation numbers, polarizations, correlations, etc., in a self--consistent way. Note in this context, that in particular magnetic ordering was investigated in a self--consistent manner.

Our numerical calculations show that ferromagnetism can be found with the theory presented in Secs. \ref{ChapTheory} and \ref{SpezLsgDerSelfen} of this paper. In addition, magnetic ordering (and thus the Curie temperature) of systems with energy gap between $f$--level and conduction band is dominated by the chosen hybridization strength. Surprisingly, the $s$--$f$--interaction strength rather acts as a destabilizing parameter. This is most likely due to our self--consistent treatment of the $f$--electron's magnetic moments, which therefore differs from conventional RKKY behavior. Contrary, fully closing the energy gap between $f$-- and $s$--band increases electronic fluctuations and, hence, destroys the ferromagnetic ordering.

As our band model shows highly interesting physical behavior, the model could be used for further research on the system's transport properties, thermodynamic quantities, or antiferromagnetic ordering properties. Another possible extension could bring different self--energies into focus, for example a $\boldsymbol{k}$--dependent one (see \cite{Henning.2012}), or a self--energy which can be obtained via moment conserving decoupling approach treatment. Also, inclusion of realistic DFT--results would improve the description of real substances, as previous works have shown (see for example \cite{Nolting.2002, Sharma.2006}).

 \section*{Acknowledgement}
The work is supported by the Deutsche Forschungsgemeinschaft (project A14 of the SFB 668).

\hrulefill

\end{document}